\documentclass{elsart}
\usepackage{graphicx}
\usepackage{epsfig}
\usepackage{amssymb}
\begin{document}
\begin{frontmatter}

\title{Radioelectric field features of extensive air showers observed
  with CODALEMA}

\author[1]{D. Ardouin}
\author[1]{A. Bell\'etoile}
\author[1]{D. Charrier}
\author[1]{R. Dallier}
\author[2]{L. Denis}
\author[3]{P. Eschstruth}
\author[1]{T. Gousset}
\author[1]{F. Haddad}
\author[1]{J. Lamblin}
\author[1]{P. Lautridou\corauthref{*}}
\ead{lautrido@in2p3.fr}
\corauth[*]{corresponding author:}
\author[4]{A. Lecacheux}
\author[3]{D. Monnier-Ragaigne}
\author[1]{O. Ravel}
\author[1]{T. Saugrin}
\author[1]{S. Valcares}
\address[1]{SUBATECH, IN2P3-CNRS, Universit\'e de Nantes, Ecole des
  Mines de Nantes, Nantes, France}
\address[2]{Station de Radioastronomie, Nan\c{c}ay, France}
\address[3]{LAL, IN2P3-CNRS, Universit\'e de Paris Sud, Orsay, France}
\address[4]{LESIA, Observatoire de Paris-CNRS UMR 8109, Meudon, France}

\begin{abstract}
Based on a new approach to the detection of radio transients
associated with extensive air showers induced by ultra high energy
cosmic rays, the experimental apparatus CODALEMA is in operation,
measuring about 1 event per day corresponding to an energy threshold
$\sim 5\times 10^{16}$~eV. Its performance makes possible for the
first time the study of radio-signal features on an event-by-event
basis. The sampling of the magnitude of the electric field along a
600~meters axis is analyzed. It shows that the electric field lateral
spread is around 250~m (FWHM). The possibility to determine with radio
both arrival directions and shower core positions is discussed.
\end{abstract}


\end{frontmatter}

\section{Introduction}

Radio emission associated with the development of Extensive Air
Showers (EAS) was investigated in the 1960's~\cite{ask,weekes}. A
flurry of experiments provided initial information about signals from
$10^{17}$~eV cosmic rays~\cite{allan}, but plagued by difficulties
(poor reproducibility, atmospheric effects, technical limitations)
efforts almost ceased in the late 1970's while ground
particle~\cite{agasa} and fluorescence~\cite{fly} detector work
continued.  With the growing interest for ultra high-energy cosmic ray
(UHECR) research involving giant surface detectors~\cite{auger}, radio
detection however appears as a promising tool for future apparatus
considering its specific advantages: low-cost, high duty cycle and
sensitivity to the longitudinal development of the showers.  Although
the rebirth of radio pulse
investigation~\cite{casa-mia,cascade-grande,gorham,sub,rice} is
recent, first available results~\cite{nature,nim-ard} demonstrate the
feasibility of EAS radio detection.

The CODALEMA (COsmic ray Detection Array with Logarithmic
Electro-Magnetic Antennas) experiment, located at the Nan\c cay radio
observatory~\cite{DAM}, is a part of this new effort. Its originality
lies in completely recording the form of the transient radio signals
together with sampling the electric field over a few hundred meters
range on the ground. This makes possible both the determination of
arrival times and directions of radio pulses and the study of their
field amplitude impact parameter dependences. The measurement of
UHECR around $10^{20}$ eV is the admitted goal of such experimental
development, but a proof-of-principle demonstration at such a large
energy would suffer from a lack of statistics without an extensive
antenna array. This point can be circumvented in a first stage by
working around $10^{17}$~eV where a measurable signal
amplitude~\cite{allan,huege} is expected at not too large distances
from the shower core. Considering a vertical shower falling upon the
detector, the predicted transient should reach 150 $\mu$V/m with a 10
ns FWHM duration  for $10^{17}$~eV cosmic
rays~\cite{allan,casa-mia,nim-ard}. Transposed in the frequency
domain, the corresponding pulse spectrum should extend from 1 to 100
MHz. Thus, with a wide band antenna it should be possible to recover
the original pulse shape, allowing for energy determination and
providing information on the nature of the primary particle with
minimal assumptions concerning its electromagnetic shower-
signature. The design of CODALEMA is based on these expectations.

CODALEMA has already provided firm evidence for a radio emission
counterpart of EAS with an estimated energy threshold of $5\times
10^{16}$~eV~\cite{nim-ard}. In the present paper, we extend the
characterization of EAS candidates giving special attention to the
electric field pattern observed on an event-by-event basis. The
experimental set-up is described in section~2 and general event
properties are given in section~3. The selection of EAS radio
candidates is discussed in section~4. Section~5 is the central part of
the present study and details first observations on EAS electric field
lateral distributions from a few illustrative examples. Section~6
discusses frequency dependences. Some conclusions are given in
section~7.

\section{The CODALEMA experiment}

\begin{figure}
 \centering
 \includegraphics[width=13cm]{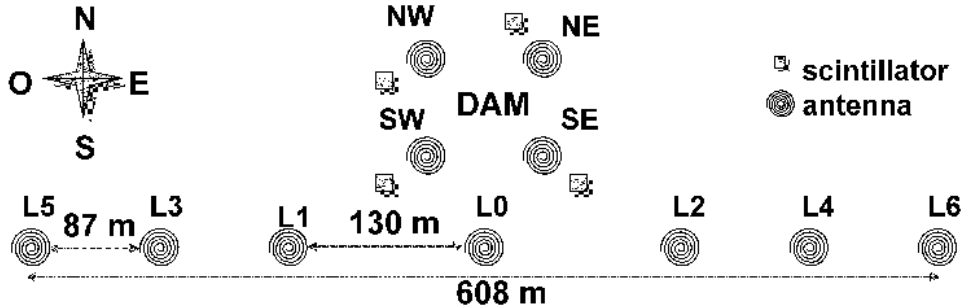}
 \caption{Current CODALEMA set-up. The particle detectors
 (scintillators) act as a trigger with a fourfold coincidence
 requirement. The 608~m long baseline in the East-West direction makes
 possible measurements of the electric field profile.}
 \label{fig:setup2}
 \end{figure}

The technical characteristics of the detector along with detection and
analysis methods have been extensively described in
Ref.~\cite{nim-ard}. As shown in Fig.~\ref{fig:setup2}, the set-up
uses 11 log-periodic antennas of the Nan\c cay Decameter
Array DAM~\cite{DAM} and 4 particle detectors originally designed as
prototypes for the Pierre Auger Observatory~\cite{boratav}.  Four of
the antennas, namely NE, SE, SW and NW, are located at the corners of
the Decameter Array (a rectangle of 87~m$\times$83~m). In order to
investigate the electric field spread, the main improvement, as
compared to the previous apparatus~\cite{nim-ard}, lies in the
instrumentation of an East-West line, 608~m long, 40~m South of the
SW-SE antenna axis.  This long baseline is equipped with 7~antennas
with a sampling interval of 87~m from L1 to L5 and L2 to L6 and 130~m
for L0, L1 and L2.

All the antennas are linked, after radio frequency (RF) signal wide
band amplification (35~dB), via low loss coaxial cables (SUHNER
S12272-04) to LeCroy digital oscilloscopes (8~bit ADC, 500~MHz
sampling frequency, 10~${\mu}$s recording time). To get enough
sensitivity to fast transients with these ADCs, the antennas are
band-pass filtered ($24-82$~MHz) so that the ADCs are not required to
handle large amplitude, low-frequency interference.
 
Each 2.3~m$^2$ particle detector module (station) has two layers of
acrylic scintillator, read out by a photomultiplier located at the
centre of each sheet. The photomultipliers have copper housings and it
has been verified~\cite{nim-ard} that no correlation exists between
individual photomultiplier signals and the presence of antenna
signals. The coincidence between the top and bottom layers is obtained
within a 60~ns time interval with a counting rate of 200~Hz per
station. The whole experiment is triggered by a fourfold coincidence
from the stations in a 600~ns time window. The corresponding counting
rate is around 0.7~events per minute.

\section{General properties of the recorded events}

\begin{figure}[h]
 \centering
 \includegraphics[width=8cm]{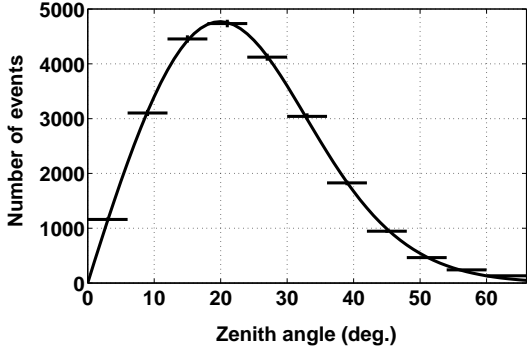}
 \caption{Cosmic ray counting rate measured during 2314 hours as a
 function of the zenith angle. The fall-off observed at low elevation
 is related to the decrease of the effective size of the array and
 detectors as well as to an increase of the detection energy threshold
 for the most inclined showers.}
 \label{fig:acceptance}
 \end{figure}

The directions of the particle showers are determined by
triangulation, using arrival times from the digitized photomultiplier
signals and assuming the particle wavefront is a plane. The angular
acceptance of the trigger device can be studied by calculating its
counting rate as a function of the zenith angle $\theta$ (see
Fig.~\ref{fig:acceptance}). A satisfactory description of the data is
obtained using a $2\pi\,\sin\theta\,\cos^2\theta$ behavior multiplied
by a Fermi-Dirac function (solid curve in
Fig.~\ref{fig:acceptance}). The latter contribution takes into account
the extinction at the detector location of large zenith angle
EAS~\cite{revenu}. Thus, the effective sky coverage extends up to
$60^\circ-70^\circ$ zenith angles. In addition, the azimuthal angle
distribution behaves as expected for our set-up. The absence of
anomalies in these distributions leads us to conclude that
satisfactory performance is achieved for triggering the radio array.

For each fourfold coincidence from the particle detectors, the 11
antenna signals are recorded. Due to the relatively low energy
threshold, only a small fraction of these air shower events is
expected to be accompanied by significant radio signals. Recognition
of radio transients is made by offline analysis using first a
37-70~MHz digital filter.  The maximum voltage is then searched for in
a 2~$\mu$s wide time window correlated to the trigger time. The
average noise and its standard deviation are calculated, for each
antenna and each event, in a 7.2~$\mu$s wide time window out of the
signal one. More details about the resulting signal determination and
the corresponding threshold criterion based on the individual event
noise can be found in Ref.~\cite{nim-ard}. When at least 3~antennas
are flagged, a triangulation procedure calculates the arrival
direction of the radio wave using a plane wavefront assumption. At
this level of selection, the counting rate of these 3-antenna events
is about 1~event every 2~hours.

\begin{figure}[t]
 \centering
 \includegraphics[width=8cm]{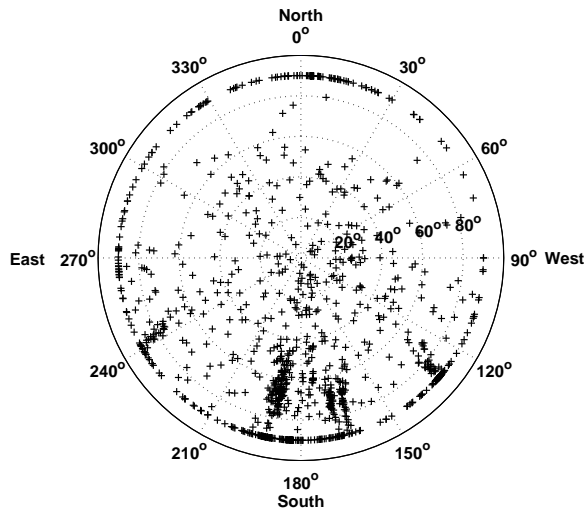}
 \caption{Sky map of the reconstructed directions of 1151 radio
 coincidences ($\ge $ 3-antennas) recorded by CODALEMA over 96~days
 and triggered by four-fold coincidences on particle detectors. Dotted
 internal circles refer to indicated values of zenith angles up to the
 horizon; the external values stand for azimuthal angles.}
 \label{fig:allevents}
 \end{figure}

The capability of the radio antenna device to reconstruct signal
directions by such triangulation is illustrated by
Fig.~\ref{fig:allevents}. It shows the distribution of the arrival
directions of events after the trigger process described above. A
substantial number of events comes from directions near the horizon
and exhibit broad accumulations at a few azimuth values. They
correspond to Radio Frequency Interference (RFI) events and the
selection discussed in the next section eliminates them. The remaining
EAS candidates turn out to be randomly distributed in the sky and have
zenith angles below 60$^\circ$ (as expected from trigger
bias). Assuming spherical wave propagation, most of the RFI were found
to originate from electrical devices located in the near environment
of the Nan\c{c}ay observatory site. Specific runs with a trigger
generated by the antennas have been performed and described in
Ref.~\cite{nim-ard}. They showed that a part of the localized sources
can emit transients at specific and restricted time sequences, in
contrast to the expected random behavior of EAS events. For example,
the group of events located around an azimuthal angle of 190$^\circ$
results from the rotating device of the primary mirror of the Nancay
Decimetric radiotelescope which generates events at well identified
time sequences.

\section{Direction and timing properties of the EAS radio events}

\begin{figure}[h]
 \centering
 \includegraphics[width=8cm]{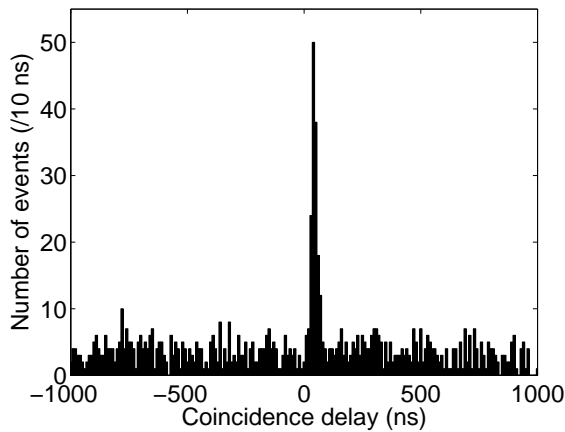}
 \caption{Distribution of time delays between the radio plane front
 and the particle plane front. The time of passage of the radio
 wavefront through a reference point, chosen as the centre of the
 triangle formed by SW, SE, L0 antennas, is compared, for all events,
 to that of the particle front. The distribution exhibits a sharp peak
 corresponding to EAS candidates.}
 \label{fig:dift1}
 \end{figure}

Our characterization of EAS radio signals starts by comparing arrival
times and incident directions as determined by the antenna array on
the one hand, and the particle detector array on the other hand. The
radio wave arrival time at any particular location is first extracted
from off-line triangulation of multi-antenna events. The corresponding
time distribution can then be compared with the particle front time
reference supplied by the scintillator signals. The time difference
distribution obtained by this procedure is shown in
Fig.~\ref{fig:dift1}. A sharp peak, a few tens of nanoseconds wide, is
clearly visible, showing an unambiguous correlation between some radio
events and particle triggers. Outside of the coincidence peak, the
flat distribution corresponds to accidental radio transients which are
not associated with air showers but which randomly occurred in the
2~$\mu$s window  where the search is conducted as described in
section~3. Being uncorrelated to
the particles, these events have a uniform arrival time
distribution. EAS candidates are those for which the arrival time
difference between the two detector systems satisfies the relation
$0\le\Delta t\le 100$~ns, corresponding to the peak in
Fig.~\ref{fig:dift1}.

\begin{figure}[h]
 \centering
 \includegraphics[width=6.5cm]{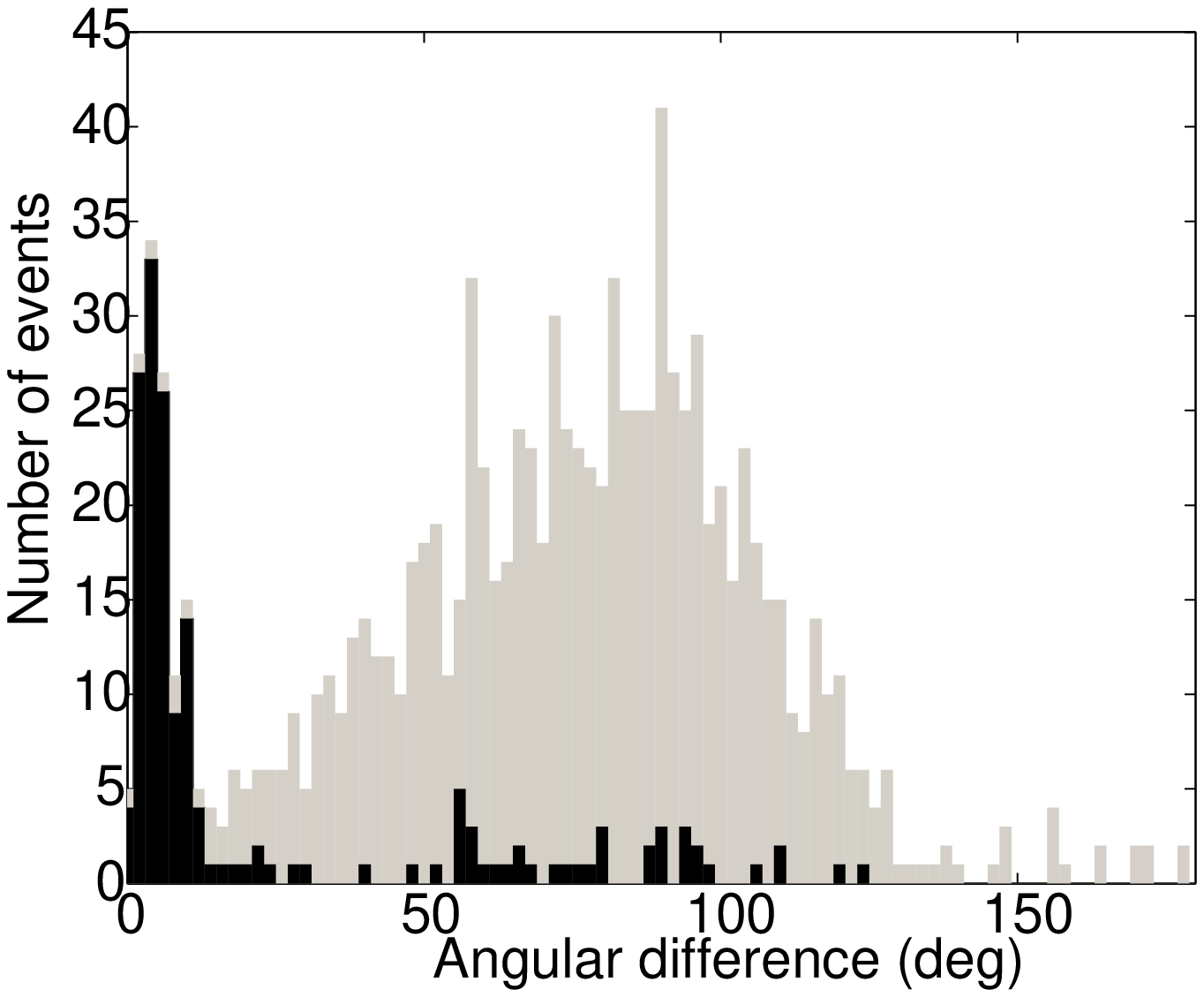}
 \includegraphics[width=6.5cm]{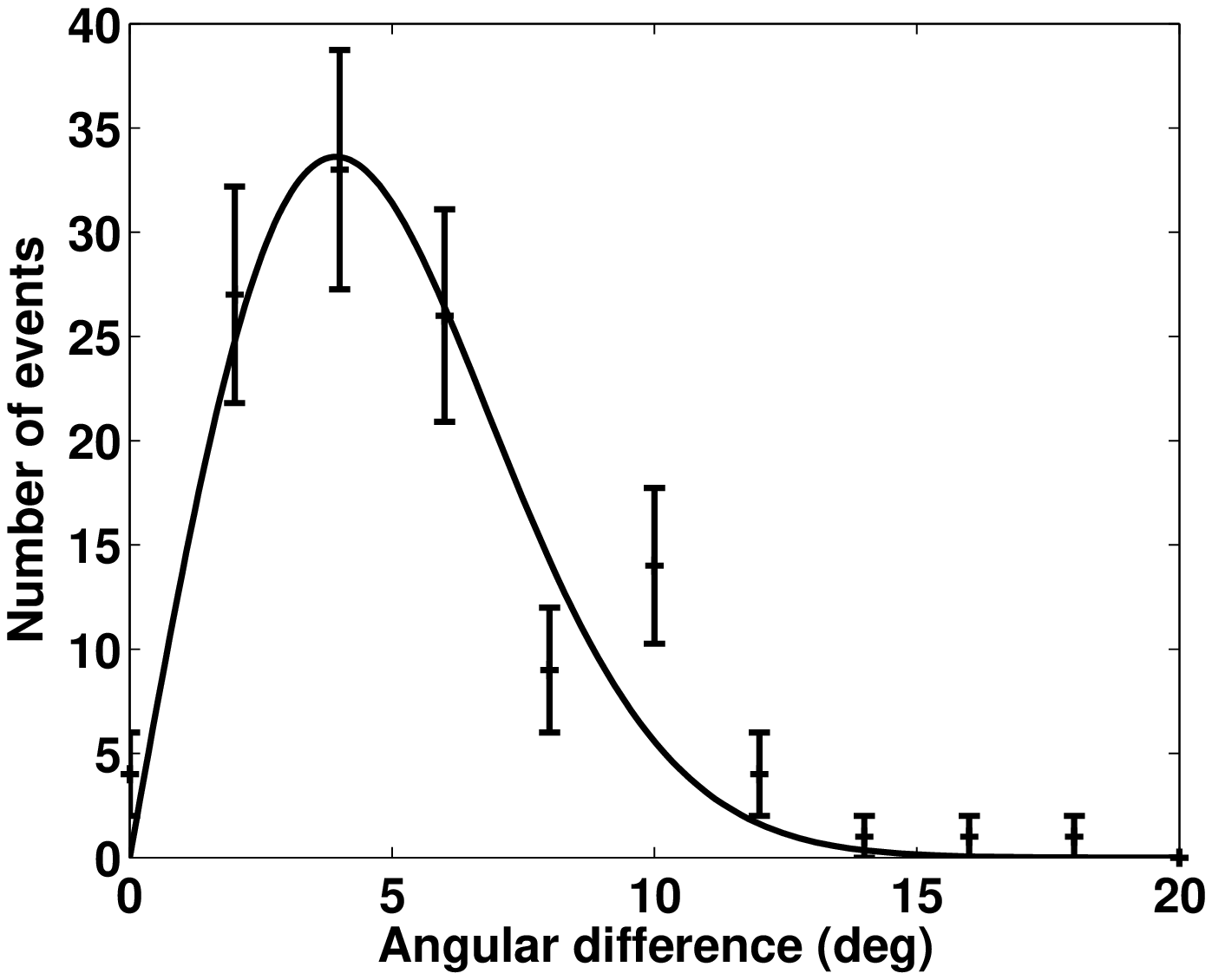}
 \caption{Left: Distribution of relative angles between particles and
 radio-pulses without time cut (grey histogram) and with a time cut
 around the coincident peak of the time difference distribution (black
 histogram). Right: Distribution of angular difference after time
 cutting around the coincident peak, fitted by an analytical
 expression of the expected angular dependence. Error bars are
 calculated as the square root of the number of events in each bin.}
 \label{fig:dift}
 \end{figure}

If the time-correlated events actually correspond to EAS, the arrival
directions reconstructed from both scintillator and antenna data
should be close to each other. Fig.~\ref{fig:dift}, left-hand side,
shows the distribution of the angle between the two reconstructed
directions, without the time cut (grey histogram) and with a time cut
of $0\le\Delta t\le 100$~ns around the peak displayed in
Fig.~\ref{fig:dift1} (black histogram). When selecting candidates in
the peak, some chance events remain and can be clearly identified by
plotting the angular difference distribution between radio pulses and
particles. Most of the chance events have a large angular difference
value and therefore true radio-particle coincidences can be selected
using a small-angle cut on this distribution. Fig.~\ref{fig:dift},
right-hand side, shows that the EAS event angular differences fit a
Gaussian distribution, centered at zero degrees and multiplied by a
sine function coming from the solid angle factor. The standard
deviation of the corresponding Gaussian is about 4~degrees. This value
combines particle detector and antenna reconstruction
accuracies. Based on the gaussian fit, the cutoff in angular
difference for true radio-particle coincidences can be set to
15$^\circ$. From the shape of the chance event distribution seen in
Fig.~\ref{fig:dift}-left between 15$^\circ$ and 100$^\circ$ the
expected number of chance events below 15~$^\circ$ is 15.

\begin{figure}[t]
 \centering
 \includegraphics[width=8cm]{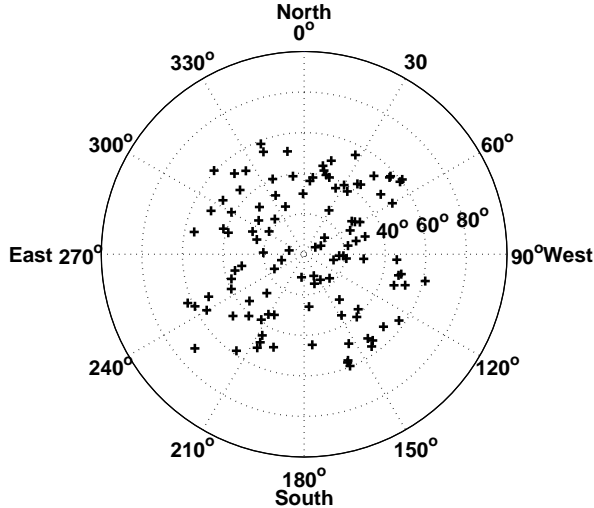}
 \caption{Sky map of the reconstructed directions of 112 air-showers
 coincidences recorded over 96~days by CODALEMA, triggered by
 four-fold coincidences on particle detectors and after time and
 angular cuts as explained in the text. Angular coordinates as in
 Fig.~\ref{fig:allevents}.}
 \label{fig:goodevents}
 \end{figure}

Recognition of chance events in the time-coincidence peak of
Fig.~\ref{fig:dift1} can be performed taking advantage of the angular
difference distribution of Fig.~\ref{fig:dift}: a third of the events
(52 out of 164) in the time-coincidence peak have angular differences
greater than 15$^\circ$ and can thus be removed. Thus, 112 events
remain as true EAS coincidences, while the chance event number (15) in
the angular range $\theta=0:15^\circ$ results in $(1/20)\times
15=0.75$ count after both time and angular coincidence cuts. The
corresponding signal directions, reconstructed as explained in
section~3, are displayed in Fig.~\ref{fig:goodevents}. The rate of
chance coincidences eliminated in this way (52 in 100~ns) is fully
compatible with the observed uniform distribution in the 2~$\mu$s
window (about 5 every 10~ns). As already mentioned, these chance
events occur mostly from directions close to the horizon, as shown in
Fig.~\ref{fig:allevents}, and are typically from RFI due to human
activities or from distant thunderstorms. As will be seen in the next
section, these events are characterized by an almost uniform electric
field amplitude, at least on the distance scale of our apparatus.

\begin{figure}
 \centering 
 \includegraphics[width=8cm]{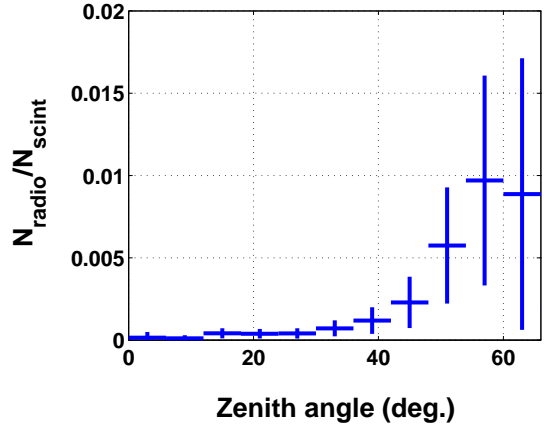} 
 \caption{Counting rate ratio of multi-antenna events to the trigger
 events versus zenith angle showing the comparative acceptance of the
 two detectors.}
 \label{fig:acceptanceratio}
 \end{figure}

In order to study the radio detection performance on confirmed EAS,
the particle detector and radio detector acceptances have been
compared as a function of the zenith angle (see
Fig.~\ref{fig:acceptanceratio}) and of the azimuthal angle. This
procedure should also bring to light possible detection inefficiency
due to antenna lobe acceptance effects. In this study, only the subset
of true radio coincidences (events present in the peak of the time
distribution satisfying the above 15$^\circ$ angular criterion) were
taken into account. Although antennas of the Decameter Array,
initially dedicated to the observation of the Sun and Jupiter, are
tilted $20^\circ$ to the south, the azimuthal distribution indicates
that the antenna array does not present a markedly higher sensitivity
for signals coming from the south. (As previously noted, the particle
detector selects EAS with zenith angle smaller than about 60$^\circ$,
see Fig.~\ref{fig:acceptance}.) As a function of zenith angle, a
higher acceptance ratio (a factor about 50 within statistics) is
observed at low elevation angles. A substantial contribution to this
evolution is coming from the fall-off of the trigger device efficiency
at large angle, due to simple geometrical effects and increase of the
energy threshold (see discussion of Fig.~\ref{fig:acceptance} in
section~3). As a matter of fact, the gross trend of the compared
acceptances can be reproduced ignoring any non trivial zenith
dependence originating from the antennas. More precisely, the antenna
gain pattern has an attenuation of less than 2~dB in the range
$\theta\le 60^\circ$. This highlights the specific and unique
advantage of a radio antenna network for the observation of inclined
showers. A more detailed and quantitative statement must take into
account several dependences such as those of the electromagnetic
signal with impact parameter~\cite{gousset}, the yield evolution in
larger atmospheric densities encountered by inclined showers and the
energy dependence of the antenna multiplicity. The scope of the next
section, which will present the analysis of field patterns, is to
bring some new valuable pieces of information regarding those
dependences.

To summarize, from 2314~hours data taking, 1151~coincidences (with at
least $\ge $ 3-antennas) were observed in the 2~$\mu$s trigger
coincidence time-window containing 112~true EAS and 52 chance
coincidence events as determined by a time cut of $0\le\Delta t\le
100$~ns around the peak of arrival time difference between the two
types of detectors and a cutoff at 15$^\circ$ in their angular
difference. After application of our selection procedure, the
resulting counting rate of EAS events with a radio signal counterpart
is close to 1~per~day. The physical characteristics of the
electromagnetic field spread associated with these radio EAS radio
events will be now considered.

\section{Electric field topologies of the EAS radio events} 

With our limited size antenna array, the number of tagged antennas per
event is highly variable, depending on the shower energy, core
position and zenith angle. Nevertheless, each tagged antenna provides
a measured voltage associated with a particular geographical location.
Thus a sampling of the electric field amplitude over the area covered
by the antenna array is possible on an event-by-event basis.

For this purpose, antenna cross-calibration has to be checked. This is
discussed in the Appendix where it is shown that antenna responses are
identical, with differences at the level of a few percent. Conversion
from ADC voltage to electric field magnitude is explained in
Ref.~\cite{nim-ard}.

\begin{figure}[h]
 \centering
 \includegraphics[width=11cm]{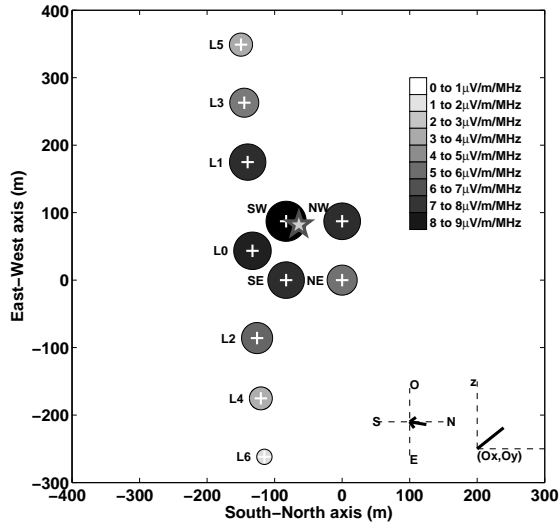}
 \caption{Footprint of a transient EAS event on the CODALEMA
 set-up. Crosses correspond to positions of each individual antenna,
 the gray-shaded circular area being proportional to the indicated
 measured electric field. The arrival direction and elevation angle of
 the event are shown in the bottom right corner. The star, near
 antenna SW for this event, indicates the shower core location
 reconstructed from the barycentre of the two distributions plotted in
 Fig.~\ref{fig:profil}.}
 \label{fig:footprint}
 \end{figure}

Fig.~\ref{fig:footprint} shows a sample EAS event with an 11-antenna
multiplicity. The signal amplitude from each antenna is represented by
the area of the gray circle. The arrival direction has been
reconstructed from both scintillator and antenna data (a difference of
1.6$^\circ$ is found between the two estimations for this event) and
indicates that it corresponds to a shower with a zenith angle of
51$^\circ$ and an azimuthal angle of 350$^\circ$.

\begin{figure}[h]
 \centering
 \includegraphics[height=5cm]{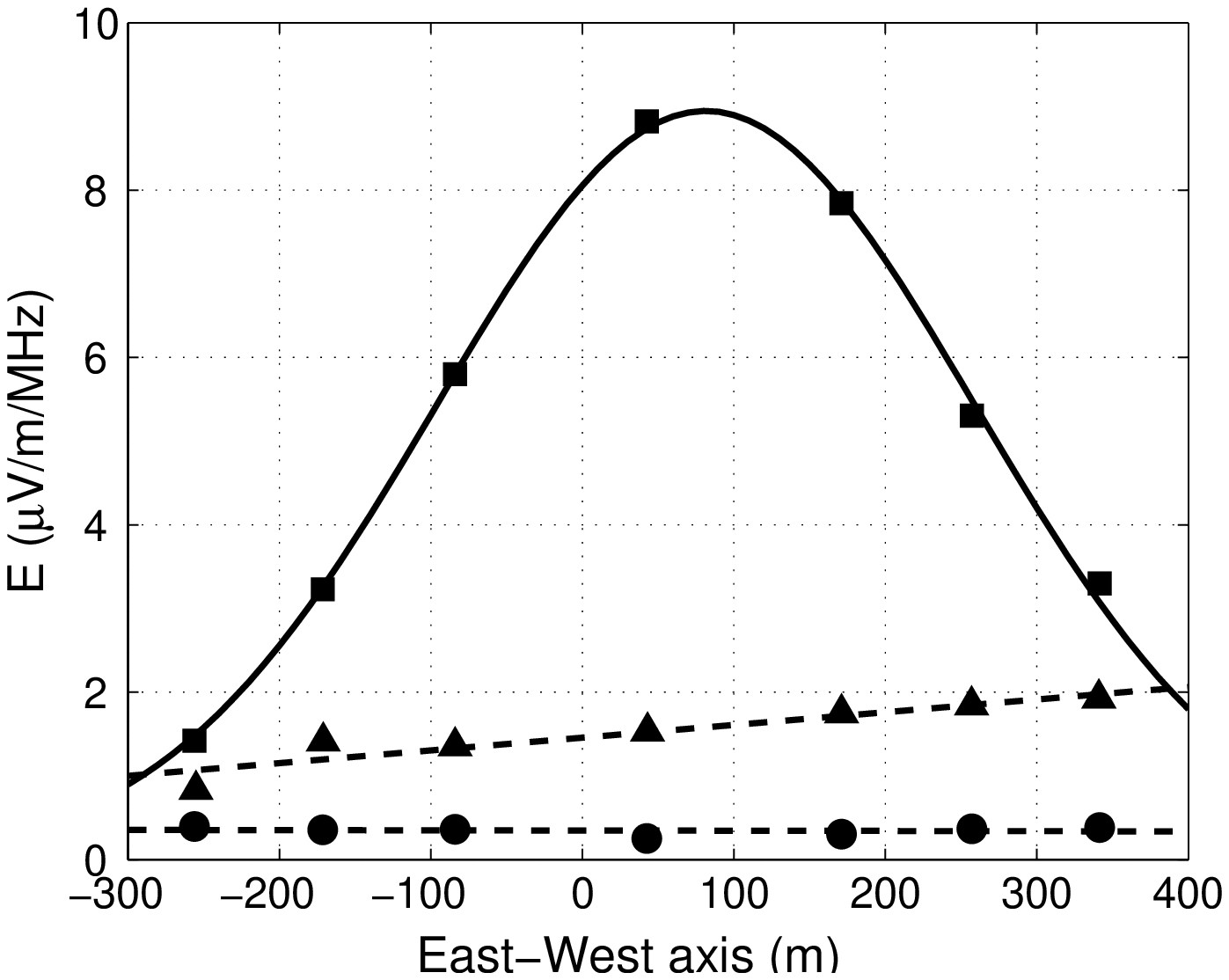}
 \includegraphics[height=5cm]{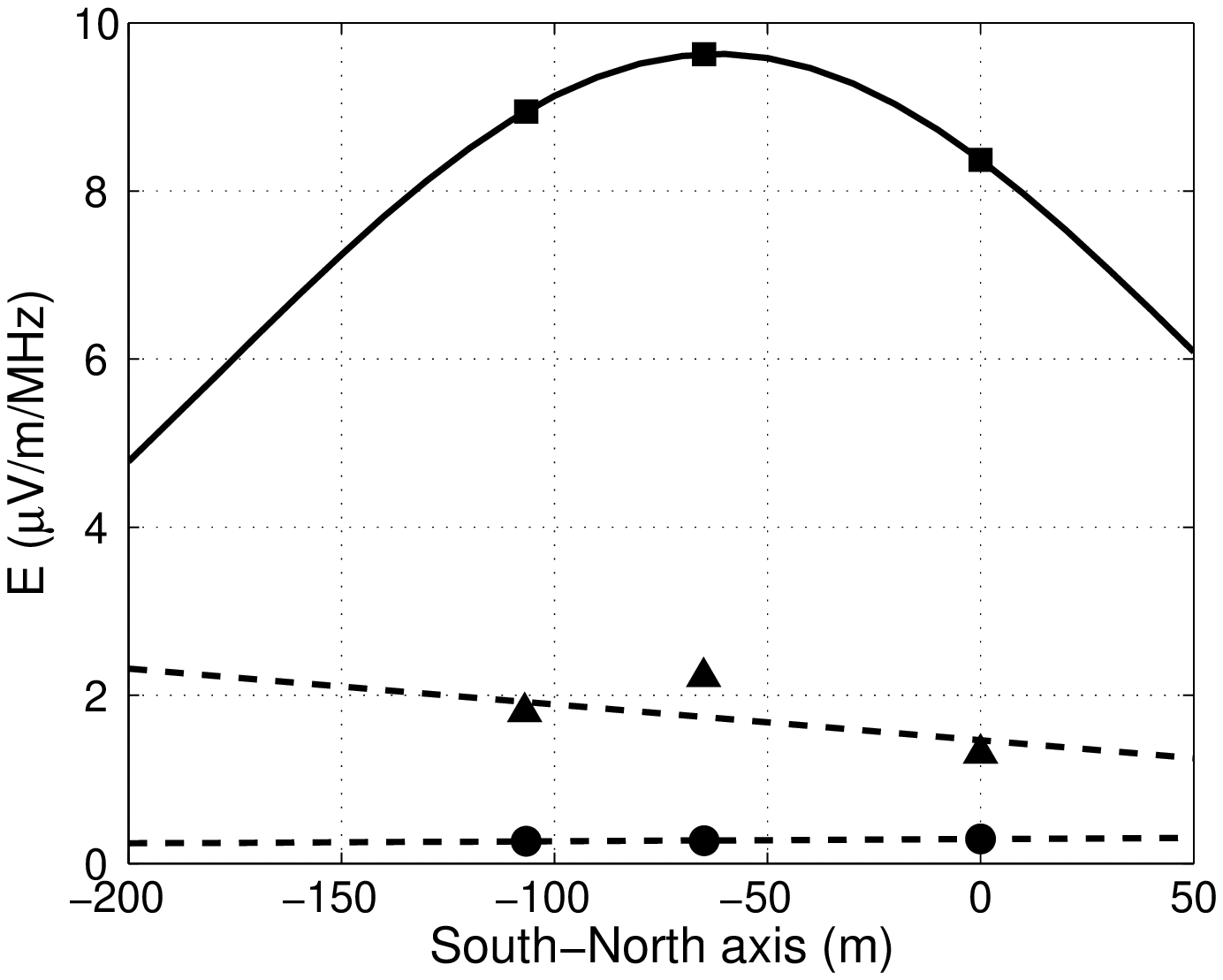}
 \caption{Electric field variations recorded on different antennas for
 an EAS event (squares, full line) and an anthropic transient
 (triangles, dashed line). Circles indicate threshold levels. Lines
 between data points are just a guide for the eye. Left: Antennas along
 the East-West line direction. Right: South-North direction (antennas
 L0,SW,NW).}
 \label{fig:profil}
 \end{figure}

The electric field distributions for this EAS event (squares) are
shown in Fig.~\ref{fig:profil}, for the East-West (left-hand side) and
South-North (right-hand side) antenna axis, together with a chance
event (triangles). The chance event has been identified using the
procedure described in section~4 and belongs to a set of events
identified as resulting from a particular RFI source (corresponding to
one of the accumulations in Fig.~\ref{fig:allevents}). Circles
indicate the threshold level as determined by the procedure mentioned
in section~3.  Fig.~\ref{fig:profil} shows that the topologies are
clearly different between EAS and RFI events. The RFI event
(triangles) displays an electric field with a quasi uniform
amplitude. This behavior corresponds to what is expected for a distant
source. To the contrary, EAS candidates falling in the vicinity of the
array should present a quite different electric field
behavior~\cite{huege}. Indeed, the EAS event (squares) shows a large
field amplitude variation depending on the position of the antenna
with respect to the shower axis. An estimation of the location of the
shower impact can be derived from the position of the barycentre. It
is indicated by a star in Fig.~\ref{fig:footprint}.

The electric field values can be compared with the galactic noise
level~\cite{allan}. Following the definition of Ref.~\cite{gorham}
\[
E^\mathrm{sky}(\mu\mathrm{V/m/MHz})=
\sqrt{20/g}\times\sqrt{\Delta t(\mu\mathrm{s})},
\ \mathrm{at}\ 55~\mathrm{MHz,}
\]
where $g$ is the antenna gain ($g=5$) and $\Delta t$ is the
integration time. The occurence of $\Delta t$ is because we define an
electric field per unit bandwidth (related to the square root of an
\emph{energy} spectral density), a quantity relevant to the
description of a finite time signal (such as EAS electric field),
whereas the sky background is a steady signal for which \emph{power}
spectral density is more appropriate. Thus, the conversion
energy$=\Delta t\times$power. For $\Delta t=1~\mu$s,
$E^\mathrm{sky}=2\mu$V$/$m$/$MHz. For pulses of duration smaller than
$1~\mu$s, as is the case of the EAS transients observed in CODALEMA,
it is possible to reduce $\Delta t$, hence increasing the
signal-to-noise ratio. In such a way, electric field as small as
$1\mu$V$/$m$/$MHz could be observed in the present analysis.

\begin{figure}[h]
 \centering
 \includegraphics[width=8cm]{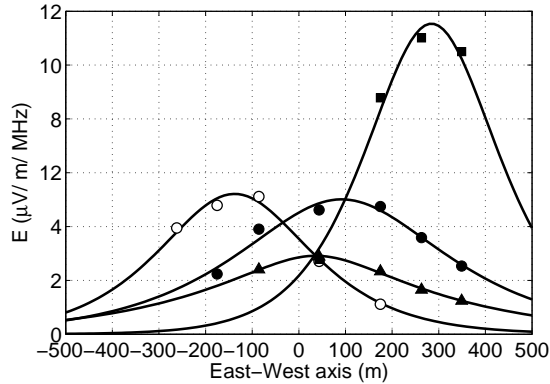}
 \caption{Electric field variations recorded on the different antennas
 in the East-West direction for four EAS events (squares). The full
 lines result from the exponential fits discussed in the text.}
 \label{fig:allprofil}
 \end{figure}

The electric field maximum in both sampling directions can only be
obtained for the subset of showers falling inside the area delimited
by the four corners of our antenna array. The present set-up is
limited in this respect due to the small extent of the array along the
South-North direction. In order to gauge the extension of the area
illuminated by the radio component of EAS, the electric field
distribution has been studied along the East-West axis without any
constraint along the South-North line. Field patterns for four typical
radio EAS events (out of 64~ with antenna multiplicity $\ge 4$) are
presented in Fig.~\ref{fig:allprofil}. EAS events show a large
variation of field amplitudes depending on the position on the E-W
axis. The differences from one event to the other are the consequence
of several geometrical (direction, impact location) and physical
(energy, depth at maximum) sources of variation in the EAS. They
cannot be disentangled using the present experimental set-up but the
patterns shown in Fig.~\ref{fig:allprofil} can be exploited by
themselves for the determination of the shower impact. Lines, drawn in
Fig.~\ref{fig:allprofil} for this purpose, are the result of a
calculation which will be explained later.

The characteristic shape of the field distribution is one genuine
signature of EAS radio events. It may help to discriminate them from
RFI events and may thus be used as an important criterion of selection
in a stand-alone radio experiment. More thorough investigations are
certainly needed to establish this possibility. The present
observations corroborate the earlier campaigns of experiments
performed in a self triggering mode~\cite{nim-ard} where the
possibility of reconstruction of event waveforms and directions has
been demonstrated. The occurrence of accidental (anthropic) close
sources of emission can be fairly well handled using specific prints
like trajectory reconstruction, field distribution or time
evolution. So far, the distributions shown in Fig.~\ref{fig:allprofil}
demonstrate that EAS electric-field measurements are feasible over
distances at least as large as 600~m and presumably up to 1~km. Such
values should be associated with the most energetic events recorded at
our site which, based on their rate, are in the range of 1--5$\times
10^{17}$~eV.

The electric field variation in the antenna-based coordinate system
depends on the arrival direction, in particular on the zenith
angle. This makes comparisons between different showers or between the
E-W and S-N axis difficult. For this purpose, it is thus preferable to
reformulate the lateral dependence of the electric field profile (EFP)
in a shower-based coordinate system, obtaining the shower axis
orientation from the reconstructed arrival direction. To carry out
this analysis an exponential profile fit $E(d)=E_0\times\exp(-d/d_0)$
has been used where $d$ is the distance between the shower axis and
each fired antenna in the event considered. This form corresponds to
the radio data fit discussed by Allan~\cite{allan}. This particular
form has a minimal number of free parameters $E_0$, $d_0$, and the E-W
and S-N coordinates of the impact point on the ground. To initiate the
4-parameter fit, the core position of the shower is first roughly
estimated by a barycentre calculation of the amplitudes on both the
South-North and East-West axis of the array.

 \begin{figure}[h]
 \centering
 \includegraphics[width=8cm]{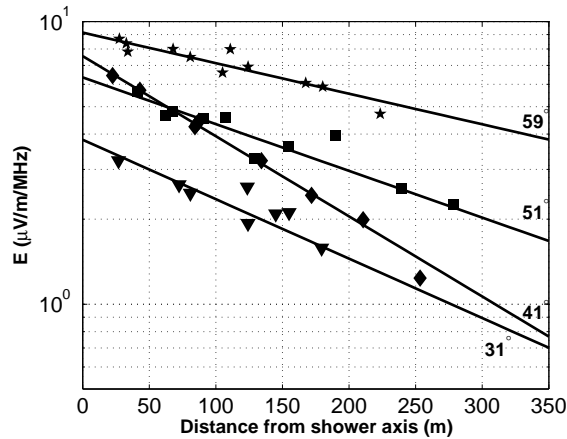}
 \caption{Electric field profiles (EFP) on a logarithmic scale for a
 set of radio EAS events recorded on CODALEMA. The measured amplitude
 in $\mu$V/m/MHz is plotted versus the distance from the antenna to
 the shower axis (in meters). The associated reconstructed zenith
 angle is indicated on each plot. Both amplitude and slope of the
 fitted exponential law are variable parameters from one event to the
 other.}
 \label{fig:LDF}
 \end{figure}

Illustrative EFPs from another sample of antenna events with
multiplicity $\ge 4$ are given in Fig.~\ref{fig:LDF}. According to the
signal threshold criteria described in section~3, for events where one
(or several) antenna is not flagged, only upper limits are used, based
on the threshold level. Data appear to be well fitted by the above
exponential law (full lines) from a few tens to several hundred
meters. As expected, the slope and amplitude parameters are variable
from one event to the other. Extrapolations of the electric field at
the locations of the shower cores could reach values as large as
10~$\mu$V/m/MHz during the data taking here discussed. Using the E-W
and S-N coordinates of the impact point thus determined as fit
parameters, the field values obtained with such fits can also be
compared to the data field values sampled in the ground coordinate
system shown in Fig.~\ref{fig:allprofil}. This is illustrated by full
lines in this plot. Data and maxima positions are very well reproduced
by this parameterization. In addition to the consistency of the
exponential law hypothesis with the shower observables, this also
demonstrates the feasibility of a shower impact determination based
only on an investigation of the electric field pattern over a limited
geometry of antennas.

 \begin{figure}[h]
 \centering
 \includegraphics[width=8cm]{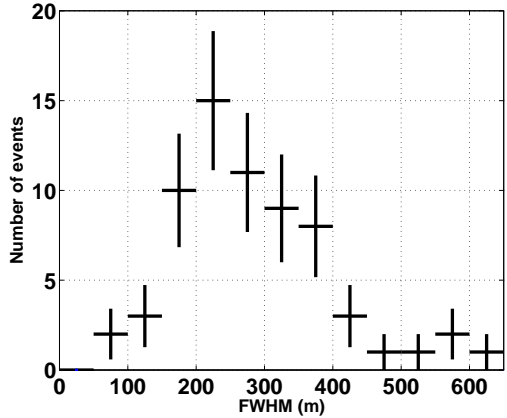}
 \caption{Distribution of the electric field FWHM extent, in meters,
  for the reconstructed events with antenna multiplicity $\ge 4$.}
 \label{fig:fwhm}
 \end{figure}

The electric field width distribution (the FWHM of the exponential) is
presented in Fig.~\ref{fig:fwhm} for the subset of $64$ events
selected from true EAS with multiplicity $\ge 4$. The mean extension
of the field is found to be around 250~m FWHM. This result may
constitute a first step toward the determination of the antenna
spacing for the design of a large radio array. In contrast to standard
measurements by ground particle detector arrays, from which a density
profile can also be extracted, we note that radio antennas are
sensitive to the overall shower development making the measurement
almost free of any particle number fluctuations.

A dependence of the electric field amplitude on the energy and the
nature of the primary particle is clearly expected~\cite{allan,huege}
but little data is available~\cite{allan,nature}. In the same way, the
slope of the exponential fit should vary with the shower zenith angle.
To our knowledge, no such simple correlations have been measured
yet. A greater amount of data is obviously needed as well as a much
larger detector array in order to go further into the physical
analysis with some statistical significance.

\begin{figure}[h]
 \centering
 \includegraphics[width=8cm]{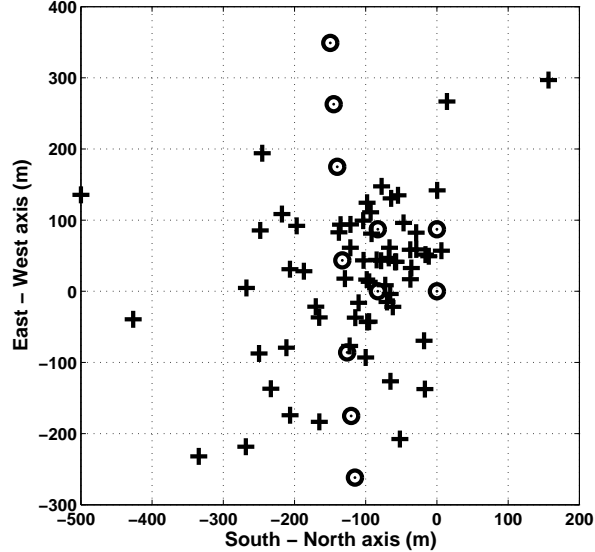}
 \caption{Core positions calculated for the complete subset of true
 EAS events with antena multiplicity $\ge 4$. Each cross corresponds
 to a reconstructed shower core position. Circles correspond to
 the positions of antennas.}
 \label{fig:core}
 \end{figure}

As long as the maximum of the field distribution is observed along one
of the sampling axis, indications of core positions can be extracted,
even if they lie outside the active area delimited by the
antennas. Core locations extracted from the exponential fits are shown
in Fig.~\ref{fig:core} for the subset of $64$ events above
defined. From the examination of the quality of the fits obtained for
the profiles (see examples in Fig.~\ref{fig:allprofil}
and~\ref{fig:LDF}), we estimate the uncertainty on the E-W position
($y_0$) around 10~m for the strongest radio events falling close to
the array centre. The uncertainty on the S-N position ($x_0$) is
somewhat larger. A variation of $y_0$ by 10~m results in a variation
of  the parameter $d_0$  by about $10\%$ to produce a fit of comparable quality. This
reflects the correlation between the different
parameters. Fig.~\ref{fig:core} shows that more core locations are
found in the vicinity of the trigger particle detector array as
expected. Far from the centre, the density of impacts is small and the
precision on core locations and  correlations with the arrangement of
antennas and particle detectors require further investigation.
The present work shows the possibility to use field profile studies for core
position determinations. Inherent limitations with the
restricted set-up geometry presently used foster the future extension
of our antenna set-up, particularly along S-N line, for core
determination purposes. An upgrade of the particle detector array will
allow useful comparison and correlation studies.

\section{Frequency dependence}

Another aspect which can be studied on an event-by-event basis is the
frequency dependence of the electric field spectrum, and taking
advantage of the size of the EW line, it is also possible to see how
the spectrum evolves with the distance of the antenna location to the
shower axis.

The general feature observed in the 30--70~MHz frequency range is a
powerlaw fall-off of the voltage with frequency. Hence significant
data for the whole frequency range are only available for the
strongest events. If in addition it is required that at least one
antenna is close ($d\le 50$~m) and one is distant ($d\ge 200$~m), we
are left with only 3 events to consider. The discussion below should
thus be considered as a foretaste of what could be achieved with a
larger statistics, or if the frequency range could be extended
downwards, below 30~MHz.

\begin{figure}[h]
 \centering
 \includegraphics[width=8cm]{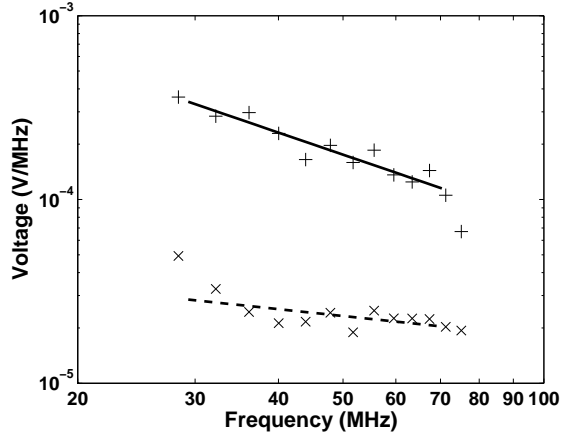}
 \caption{Voltage spectra and simple least square fits of both the
 signal ($+$) and the background ($\times$).}
 \label{fig:spectra}
 \end{figure}

The analysis consists in selecting, for each antenna signal in a given
event, a window of 256 points surrounding the radio pulse (time
signal) and 16 distinct windows of 256 points each outside the signal
window (background). Then, Fourier transforms of both sets are
calculated in order to select those pulses that have their spectra
well above the noise spectrum. The latter is estimated as the average
spectral amplitude over the 16 background
windows. Fig.~\ref{fig:spectra} shows an example of such a
comparison. The fall-off in the range 30--70~MHz can be reasonably
well described by a power-law $V_\nu=K\times\nu^{-a}$ in which $a$ can
be estimated on each signal. For this purpose, only frequencies in the
range 30--70~MHz are retained under the condition that the signal
amplitude spectrum exceeds four times that of the background. We then
performed a least square log-log fit of the signals having at least 10
frequency points fulfilling the above criterion.

\begin{figure}[h] 
 \centering
 \includegraphics[width=8cm]{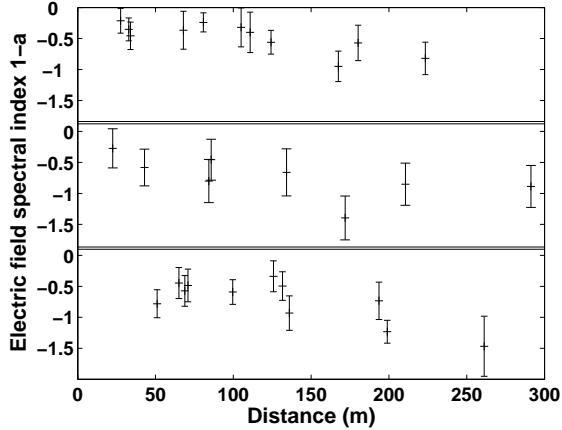}
 \caption{$1-a$ vs shower-antenna distance (in meters) for three
 events. The three events shown are those which have both a close,
 $d\le 50$~m, and a distant antenna, $d\ge 200$~m.}
 \label{fig:avsd}
 \end{figure}

An example of spectral index dependence on distance is shown in
Fig.~\ref{fig:avsd} for the electric field ($E_\nu\propto\nu\times
V_\nu$). For the small event set where such a study could be
performed, we found good descriptions of the signal spectra with the
power-law fall-off $E_\nu=K\times\nu^{1-a}$ with $1-a$ ranging from
$-1.5$ to sligthly below 0. Confidence interval on $a$ is about
$0.2$. The variation with distance is only mild if any.

The data collected is too sparse to draw any conclusion. From the
discussion in Ref.~\cite{allan}, the frequency spectrum should be
constant in the MHz range and fall-off at larger frequencies as a
consequence of the loss of coherence. The frequency cut-off and
precise behaviour of the fall-off depend on the geometrical and
physical characteristics of the shower and on the antenna
location. For example, the time duration of the electric field pulse
is expected to rise, and consequently the frequency cut-off is
expected to decrease, when the impact parameter increases~\cite{allan,nim-ard}. It follows
that a constant spectral index is not expected in the full frequency
range but rather it should evolve from $1-a=0$ at small frequency to
negative values beyond the frequency cut-off. The observed constancy
of $1-a$ may then just reflect the too-limited accessible frequency
range of the present study. The observed spectral index variation from
one signal to the next may be attributed to the change of the
coherence condition, in particular the frequency cut-off. Since the
latter depends on the impact parameter~\cite{allan}, a more thorough
study would be of great interest to complement the impact parameter
evolution described in section~5.

\section{Summary and outlook}

Features of the electric field transients generated by more than a
hundred extensive air shower events have been observed with
CODALEMA. Through the evolution of this sample of events, the observed
characteristics allow the determination of the EAS core location
together with the electric field magnitude and spread, on an
event-by-event basis using an electric field profile
function. Electric field profiles show slopes and amplitudes which are
variable between events. These patterns offer possibilities to
discriminate between EAS events and radio frequency
interferences. Frequency dependence of EAS electric field has been
also observed. Such detailed electric field correlations or
characteristics have been predicted to be related to important
physical quantities such as the EAS energy, the nature of their
primaries and various important shower evolution
parameters~\cite{allan,huege}. The present results show for the first
time the feasability of such studies on an event-by-event basis using
a radio detection set-up in combination with particle detectors. More
complete works and higher statistics are clearly necessary to
establish the exact nature of the physical correlations detectable by
the measurements of the radioelectric field features.

In addition, the various sources that can influence the electric field
distribution in a given event are hard to disentangle with the limited
set-up described here. An upgrade with new particle detectors is
already underway to provide more information on each event. They will
give an independent determination of the core position and an estimate
of the shower energy, invaluable information at the present stage of
development of the radio technique.

It has been emphasized that the observed field pattern on the various
antennas constitutes a clear ``radio'' signature. This suggests that
it may be possible to discriminate an EAS event from a fortuitous one,
using a self triggering array of radio antennas. This is one further
step toward a stand-alone system that could be deployed over a large
area or added to an existing surface detector such as the Pierre Auger
Observatory. The radio signals could then provide complementary
information about the longitudinal development of the shower, as well
as the ability to lower the energy threshold. More data and technical
upgrading are planned in order to examine the contribution that the
radio detection method could bring to the determination of the energy
and nature of ultra high energy cosmic rays.

The fact that inclined or horizontal radio waves could be efficiently
detected with our set-up may be of great interest to EAS detection if
the electric field patterns discussed in the present study turn out to
allow for a clear discrimination between EAS radio events and other
noise transients close to the horizon. Among various particles able to
generate air showers, neutrinos are thought to probably be the only
ones which could address the nature and source location of ultra
energetic cosmic rays. Taking into account the sensitivity of the
radio detection method to inclined extensive air
showers~\cite{gousset}, the presented results could considerably
enlarge the scope of such studies for cosmology related phenomena.

\appendix

\section{Antenna response comparison}

Comparison of the antenna and associated electronics responses is done
using the diurnal behavior of the ambient electric field in the
frequency band of interest. The bulk of the radio data, which does not
contain any transients, shows a steady variation of the electric field
during the day, at the level of $\pm 10\%$, with identical time
evolution from one antenna to the other. This electric field component
can be considered as uniform over the antenna array and used as a
common reference to check for antenna gain line up. The observed
deviation is smaller than $0.5$~dB.

\begin{figure}
\centering
\includegraphics[width=8cm]{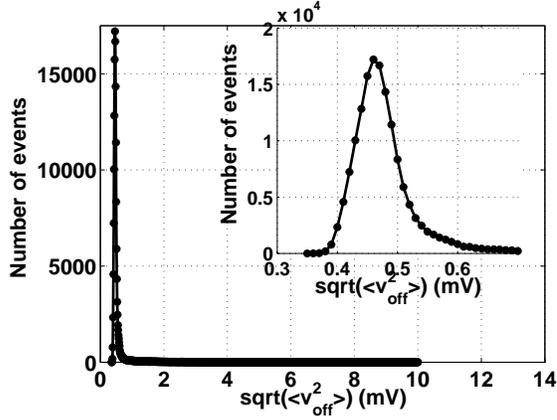}
\caption{Distribution of $\sqrt{\mu_{n}}$ on antenna L1 over the full
  scale in mV. An accumulation is visible at small values. Higher
  excursions are due to occasional pollution incidents in the
  frequency band. The inset focuses on the peak that is of interest in
  our case as it corresponds to the absence of disturbances.}
\label{fig:noisedist}
\end{figure}

For each event, the time averaged power $\mu_{n\,j}=\langle
P\rangle_{\mathrm{noise}\,j}$ is calculated on each antenna $j$ in the
noise window of the 37-70~MHz filtered signal. If this component of
the signal originates from a common source which distributes it over
the whole antenna array, $\mu_{n\,j}$ will reflect the gain of the
channel. Fig.~\ref{fig:noisedist} shows the distribution of $\mu_n$
for one antenna over 8 months. The distribution extends to large
values attributed to sources emitting occasionally or intense solar
phenomena. The peak at small values, displayed on a smaller scale in
the inset, corresponds to background conditions where no coherent
signal has been seen in the frequency spectra. In order to remove the
large fluctuations, a cut is applied on the data to deal only with the
subset of events belonging to the peak.

\begin{figure}
\centering
\includegraphics[width=8cm]{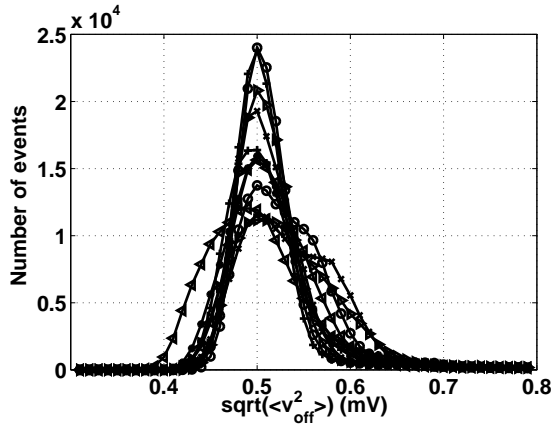}
\caption{Distribution of $\sqrt{\mu_{n\,j}}$ in mV for the whole
  array.}
\label{fig:crosscalibration}
\end{figure}

For every antenna, when plotting $\mu_{n}$ versus time over several
months, the general trend of the variation from one event to the next
is an oscillation with a period of one day. No correlation with the
surrounding human activity is observed. This pattern, seen on every
antenna with identical phased periodicity and deviation, should be
associated to a single origin which can be considered as a source
shining uniformly over the antenna array. Thus, having identified a
common reference, relative gains can be compared.  This is shown in
Fig.~\ref{fig:crosscalibration} for the entire antenna array.

Calibration of each antenna using strong celestial radiosources is
possible via interferometry using the full Decameter Array as a
reference detector. This procedure is underway.

\end{document}